\documentclass[aps,floats,superscriptaddress,floatfix,nofootinbib]{revtex4}
\usepackage{graphicx}
\usepackage{color,lscape,fancybox,pifont}
\usepackage{wasysym}

\begin{document}

\title{Defining and identifying communities in networks}

\author{Filippo Radicchi}
\affiliation{Dipartimento di Fisica, Universit\`a di Roma ``Tor Vergata'',
Via della Ricerca Scientifica 1, 00133 Roma, Italy} 
\author{Claudio Castellano}
\affiliation{Dipartimento di Fisica, Universit\`a di Roma ``La Sapienza''
and INFM-SMC, Unit\`a di Roma 1, P.le A. Moro 5, 00185 Roma, Italy}
\author{Federico Cecconi}
\affiliation{Institute of Cognitive Sciences and Technologies C.N.R.,
Viale Marx, 15, 00137, Roma, Italy}
\author{Vittorio Loreto}
\email{loreto@roma1.infn.it}
\affiliation{Dipartimento di Fisica, Universit\`a di Roma ``La Sapienza''
and INFM-SMC, Unit\`a di Roma 1, P.le A. Moro 5, 00185 Roma, Italy}
\author{Domenico Parisi}
\affiliation{Institute of Cognitive Sciences and Technologies C.N.R.,
Viale Marx, 15, 00137, Roma, Italy}

\begin{abstract}

The investigation of community structures in networks is an important
issue in many domains and disciplines.  This problem is relevant for
social tasks (objective analysis of relationships on the web),
biological inquiries (functional studies in metabolic and
protein networks) or technological problems (optimization of large
infrastructures).  Several types of algorithms exist for revealing the
community structure in networks, but a general and quantitative
definition of community is not implemented in the algorithms, leading
to an intrinsic difficulty in the interpretation of the results
without any additional non-topological information. In this paper we
deal with this problem by showing how quantitative definitions of
community are implemented in practice in the existing algorithms.
In this way the algorithms for the
identification of the community structure become fully self-contained.
Furthermore, we propose a new local algorithm to detect communities
which outperforms the existing algorithms with respect to
computational cost, keeping the same level of reliability.  The new
algorithm is tested on artificial and real-world graphs.  In
particular we show how the new algorithm applies to a network of
scientific collaborations, which, for its size, can not be attacked
with the usual methods. This new type of local algorithm could open
the way to applications to large-scale technological and biological
applications.

\end{abstract}

\maketitle

\section{Introduction}
Evidence has rapidly grown in the past few years that very diverse
systems in many different fields can be described as complex networks,
i. e. assemblies of nodes and edges with nontrivial topological
properties~\cite{Barabasi02, Newman03}.  Examples range from
technological systems (the Internet and the web~\cite{AJB99,Broder00})
to biological (epidemiology~\cite{MN00,PV01}, metabolic
networks~\cite{Jeong00,WF01,barabasi_mod}, ecological
webs~\cite{Montoya02,DWM02,CGA02,gcalda}) and social
systems~\cite{Redner,Newman01bis} (scientific collaborations,
structure of large organizations).

In this paper we deal with a topological property of networks, the
community structure, that has attracted a great deal of interest very
recently.  The concept of community is very common and it is linked to
the classification of objects in categories for the sake of
memorization or retrieval of information.  From this point of view the
notion of community is very general and, depending on the context, can
be synonymous of module, class, group, cluster, etc. Among the many
contexts where this notion is relevant it is worth mentioning the
problem of modularity in metabolic or cellular
networks~\cite{barabasi_mod,rives} or the problem of the
identification of communities in the web~\cite{webcomm}.  This last
issue is relevant for the implementation of search engines of new
generation, content filtering, automatic classification or the
automatic realization of ontologies.

Given the relevance of the problem it is crucial to construct
efficient procedures and algorithms for the identification of the
community structure in a generic network. This, however, is a highly
nontrivial task.

Qualitatively, a community is defined as a subset of nodes within the
graph such that connections between the nodes are denser than
connections with the rest of the network. The detection of the
community structure in a network is generally intended as a procedure
for mapping the network into a tree (Fig.~\ref{Fig1}). In this tree
(called dendrogram in social sciences) the leaves are the nodes while
the branches join nodes or (at higher level) groups of nodes, thus
identifying a hierarchical structure of communities nested within each
other.

Several algorithms to perform this mapping are known in the
literature.  The traditional method is the so-called hierarchical
clustering~\cite{Wassermanbook}. For every pair $i,j$ of nodes in the
network one calculates a weight $W_{i,j}$, which measures how closely
connected the vertices are. Starting from the set of all nodes and no
edges, links are iteratively added between pairs of nodes in order of
decreasing weight. In this way nodes are grouped into larger and
larger communities and the tree is built up to the root, which
represents the whole network. Algorithms of this kind are called
agglomerative.

For the other class of algorithms, called divisive, the order of
construction of the tree is reversed: one starts with the whole graph
and iteratively cuts the edges, thus dividing the network
progressively into smaller and smaller disconnected sub-networks
identified as the communities.  The crucial point in a divisive
algorithm is the selection of the edges to be cut, which have to be
those connecting communities and not those within them.  Very
recently, Girvan and Newman (GN) have introduced a divisive algorithm
where the selection of the edges to be cut is based on the value of
their ``edge betweenness''~\cite{Girvan02}, a generalization of the
centrality betweenness introduced by Anthonisse~\cite{Anthonisse71}
and Freeman~\cite{Freeman77}.  Consider the shortest paths between all
pairs of nodes in a network.  The betweenness of an edge is the number
of these paths running through it. It is clear that when a graph is
made of tightly bound clusters, loosely interconnected, all shortest
paths between nodes in different clusters have to go through the few
inter-clusters connections, which therefore have a large betweenness
value.  The single step of the GN detection algorithm consists in the
computation of the edge betweenness for all edges in the graph and in
the removal of those with the highest score. The iteration of this
procedure leads to the splitting of the network into disconnected
subgraphs that in their turn undergo the same procedure, until the
whole graph is divided in a set of isolated nodes.  In this way the
dendrogram is built, from the root to the leaves.

The GN algorithm represents a major step forward for the detection of
communities in networks, since it avoids many of the shortcomings of
traditional methods~\cite{Girvan02,Newman_cond-mat}. This explains why
it has been quickly adopted in the past year as a sort of standard for
the analysis of community structure in
networks~\cite{holme,Wilkinson02,gleiser,guimera,tyler}.

This paper follows a different track by proposing an alternative
strategy for the identification of the community structure. This
complementary approach follows from the need of addressing the two
following issues.

\begin{enumerate}
\item
In general algorithms define communities operationally as what the
they finds. A dendrogram, i. e. a community structure, is always
produced by the algorithms down to the level of single nodes,
independently from the type of graph analyzed. This is due to the lack
of explicit prescriptions to discriminate between networks that are
actually endowed with a community structure and those that are not.
As a consequence, in practical applications one needs additional, non
topological, information on the nature of the network to understand
which of the branches of the tree have a real significance. Without
such information it is not clear at all whether the identification of
a community is reliable or not. There have been two noticeable
proposals to solve this problem. In particular it is worth mentioning
the approach proposed by Wilkinson and Huberman~\cite{Wilkinson02},
which is limited to the lowest level of the community structure and
specific to algorithms based on betweenness. More
recently~\cite{Newman_cond-mat}, Newman and Girvan have introduced an
{\em a posteriori} measure of the strength of the community structure,
which they called modularity. More precisely, the modularity estimates
the fraction of inward links in a community minus the expectation
value of the same quantity in a network with the same community
divisions but random connections between the nodes. This quantity
definitely gives an indication of the strength of the community
structure, even though the lack of the implementation of a
quantitative definition of community does not allow to discriminate in
an objective way meaningful communities.

\item
The ``edge betweenness algorithm'' is computationally costly, as
already remarked by Girvan and
Newman~\cite{Girvan02,Newman_cond-mat}. Evaluating the score for all
edges requires a time of the order of $M N$, where $M$ is the number
of edges and $N$ the number of nodes. The iteration of the procedure
for all $M$ edges leads in the worst case to a total scaling of the
computational time as $M^2 N$, which makes the analysis practically
unfeasible already for moderately large networks (of the order of
$N=10000$~\cite{Newman_cond-mat}).

\end{enumerate}

In this paper we propose solutions to both these problems.  First we
introduce a general criterion for deciding which of the subgraphs
singled out by the detection algorithms are actual communities. We
discuss in detail the case of two quantitative definitions of
community.  In this way we transform the GN algorithm in a
self-contained tool.  Secondly we present an alternative algorithm,
based on the computation of local quantities, which gives, in
controlled cases, results of accuracy comparable to the GN method,
while largely outperforming it from the point of view of the
computational speed.\footnote{It is worth mentioning that after the
completion of this work Newman has proposed a new agglomerative
algorithm to address the issue of the computational
efficiency~\cite{newman_fast}.}

The outline of the paper is as follows.  In section II we introduce
the definitions of community.  In section III we show how these
definitions can be implemented in a generic divisive algorithm in
order to make it self-contained and we present tests on some
computer-generated and real networks. In section IV we present a new
and fast algorithm for the detection of communities and we compare its
performance with the GN algorithm. Section V is devoted to the
application of the new algorithm to a network of scientific
collaborations  which, because of its size, is hard to analyze
with the Girvan-Newman method. We finally draw some
conclusions and discuss the perspectives of this work.

\section{Quantitative definitions of community}

The idea to solve the first of the problems discussed above is very
simple: the algorithm that builds the tree just selects subgraphs
that are candidate to be considered communities.
One has then to check whether they are actually such by using a
precise definition. If the subgraph does not meet the criterion,
the subgraph isolated from the network is not a community and the
corresponding branch in the dendrogram should not be drawn.

As mentioned above, a community is generally thought as a part of a
network where internal connections are denser than external ones. To
sharpen the use of detection algorithms a more precise definition is
needed. Many possible definitions of communities exist in the
literature~\cite{Wassermanbook}. Here we consider explicitly the
implementation in the algorithms of two plausible definitions of
community which translate into formulas the sentence above.

The basic quantity to consider is $k_i$, the degree of a generic node
$i$, which in terms of the adjacency matrix $A_{i,j}$ of the network
$G$ is $k_i = \sum_j A_{i,j}$~\footnote{The adjacency matrix fully
specifies the topology of the network. In the simplest case of an
unweighted, undirected network, it is equal to 1 if $i$ and $j$ are
directly connected, zero otherwise.}. If we consider a subgraph $V
\subset G$, to which node $i$ belongs, we can split the total degree
in two contributions: $k_i(V) = k_i^{in}(V) + k_i^{out}(V)$.
$k_i^{in}(V) = \sum_{j \in V} A_{i,j}$ is the number of edges
connecting node $i$ to other nodes belonging to $V$.  $k_i^{out}(V)=
\sum_{j \notin V} A_{i,j}$ is clearly the number of connections toward
nodes in the rest of the network.

\vspace{0.3cm}
\textbf{Definition of community in a strong sense}
\vspace{0.3cm}

The subgraph $V$ is a community in a strong sense if

\begin{equation}
 k_{i}^{in}(V) > k_{i}^{out}(V) , \;\; \forall i \in V.
\label{strong}
\end{equation}

In a \textit{strong} community each node has more connections within
the community than with the rest of the graph. This definition coincides
with the  one proposed in~\cite{webcomm} in the framework of the
identification of web communities.

\vspace{0.3cm}
\textbf{Definition of community in a weak sense}
\vspace{0.3cm}

The subgraph $V$ is a community in a weak sense if
\begin{equation}
\sum_{i \in V} k_{i}^{in}(V) > \sum_{i \in V} k_{i}^{out}(V).
\label{weak}
\end{equation}

In a \textit{weak} community the sum of all degrees within $V$ is
larger than the sum of all degrees toward the rest of the network.

Clearly a community in a strong sense is also a community in a weak
sense, while the converse is not true.  

It is worth mentioning that our definitions of community, though very
natural, do not represent the only possible choice. Several other
possible definitions, possibly more appropriate in some cases, exist
and are described in~\cite{Wassermanbook}. Among them for instance the
definition of the so-called $LS-set$ goes in the direction of our
strong definition even though extremely more stringent. An $LS-set$ is
a set of nodes such that {\bf each} of its proper subsets has more
ties to its complement within the set than outside. On the other hand
the definition of $k$-core is roughly, although not exactly,
equivalent to our weak definition. A $k$-core is defined as a subgraph
in which each node is adjacent to at least a minimum number, $k$, of
the other nodes in the subgraph.

\section{Self-contained algorithms}

From the definitions given above, it is apparent that, if a network is
randomly split in two parts, one very large and the other with only
few nodes, the very large part almost always fulfills the definition
of community.  In order to deal with this problem, let us consider the
Erd\"os-Renyi random graph~\cite{Bollobasbook}. If we cut at random
the graph in two parts containing $\alpha N$ and $(1-\alpha) N$ nodes,
respectively, it is easy to evaluate analytically the probability
$P(\alpha)$ that the subgraph containing $\alpha N$ nodes fulfills the
weak or the strong definition (details will be given in a forthcoming
publication). It turns out that, as
soon as $N$ is sufficiently large, the probability is very close to a
step function around $\alpha=0.5$. Hence it is extremely likely that,
in a random graph randomly cut in two parts, the largest one is a
community according to the previous definitions. However it is
extremely unlikely that both subgraphs fulfill simultaneously the
definitions: therefore if we accept divisions only if both groups
fulfill the definition of community, we correctly find that a random
graph has no community structure.  We extend this criterion to the
general case: if less than two subgraphs obtained from the cut satisfy
the definitions, then the splitting is considered to be an artifact
and disregarded.

We can now summarize the improved self-contained version of the GN algorithm.
\begin{enumerate}
\item Choose a definition of community;
\item Compute the edge betweenness for all edges and remove those
with the highest score;
\item If the removal does not split the (sub-)graph go to point 2;
\item If the removal splits the (sub-)graph, test if at least two of
the resulting subgraphs fulfill the definition.  If they do, draw the
corresponding part of the dendrogram;
\item Iterate the procedure (going back to point 2) for all the
sub-graphs until no edges are left in the network.
\end{enumerate}
It is important to remark that the quantities appearing in
Eqs.~(\ref{strong}) and ~(\ref{weak}) must always be evaluated with
respect to the full adjacency matrix.  The application of this
procedure to a network produces a tree, where every branch splitting
represents a meaningful (with respect to the definition) separation in
communities.

It is now possible to blindly test the effectiveness of the GN
algorithm. We have considered the ``artificial'' graph already
discussed by Girvan and Newman. It is a simple network with $N$ nodes
divided into four groups: connections between pairs within a group are
present with probability $p_{in}$, while pairs of nodes in different
groups are connected with probability $p_{out}$. As the probability
$p_{out}$ grows from zero, the community structure in the network
becomes less well defined.

For every realization of the artificial graph the application of the
detection algorithm generates a tree. We consider the algorithm to be
successful if the four communities are detected, each node is
classified in the right community and the communities are not further
subdivided.

Fig.~\ref{Fig4} presents the comparison of the fraction of successes
for the modified GN algorithm with the expected value computed
analytically. We see that the GN algorithm captures 
very well the existence of communities in a strong sense, while it
performs less well for the weak definition.  However, one should not
be misguided by the quantity presented in Fig.~\ref{Fig4}. By looking
at a softer measure of success, the fraction $f$ of nodes not
correctly classified, one realizes that, when the algorithm with weak
definition seems to fail, it correctly identifies the four communities
and it misclassifies only a few nodes up to much higher values of
$p_{out}$.  The deviations from the theoretical behaviour observed for
small values of $p_{out}$ are due to the possibility that one or more
of the four communities are further split in smaller
sub-communities. This event, not taken into account in the analytical
calculation, becomes very unlikely as the size of the system
increases.

\section{A new fast algorithm}

The Girvan-Newman algorithm is computationally expensive because it
requires the repeated evaluation, for each edge in the system,
of a global quantity, the betweenness, whose value depends on the properties
of the whole system.  Despite smart methods to compute the edge betweenness
simultaneously for all edges~\cite{Newman01,Brandes01}, the evaluation of such
quantity is the time consuming part of the procedure. As a consequence
the time to analyze completely a network turns out to grow fast with
its size, making the analysis unfeasible for networks of
size larger than around $10000$ nodes~\cite{Newman_cond-mat}.

In order to overcome this problem we introduce a new kind
of divisive algorithm which requires the consideration of
local quantities only, and is therefore much faster than the GN algorithm.
The fundamental ingredient of a divisive algorithm is a
quantity which can single out edges connecting nodes belonging to different
communities. We consider the {\it edge clustering coefficient}, defined,
in analogy with the usual node clustering coefficient, as the number
of triangles to which a given edge belongs, divided by the
number of triangles that might potentially include it, given the
degrees of the adjacent nodes.  More formally, for the edge connecting
node $i$ to node $j$, the edge clustering coefficient is
\begin{equation}
C_{i,j}^{(3)}= \frac{z_{i,j}^{(3)}}{\min[(k_i-1),(k_j-1)]},
\label{edgeclust}
\end{equation}
where $z_{i,j}^{(3)}$ is the number of triangles built on that edge and
$\min[(k_i-1),(k_j-1)]$ is the maximal possible number of them.

The idea behind the use of this quantity in a divisive algorithm
is that edges connecting nodes in different communities
are included in few or no triangles,
and tend to have small values of $C_{i,j}^{(3)}$.  On
the other hand many triangles exist within clusters. Hence the
coefficient $C_{i,j}^{(3)}$ is a measure of how inter-communitarian
a link is.
A problem arises when the number of triangles is zero, because
$C_{i,j}^{(3)}=0$, irrespective of $k_i$ and $k_j$, (or even
$C_{i,j}^{(3)}$ is indeterminate, when $\min[(k_i-1),(k_j-1)]=0$).
To remove this degeneracy we consider a slightly modified quantity by
using, at the numerator, the number of triangles plus one:
\begin{equation}
{\tilde C}_{i,j}^{(3)}= \frac{z_{i,j}^{(3)}+1}{\min[(k_i-1),(k_j-1)]}.
\end{equation}

By considering higher order cycles we can define, in much the same
way, coefficients of order $g$ as:
\begin{equation}
{\tilde C_{i,j}}^{(g)}= {z_{i,j}^{(g)}+1 \over s_{i,j}^{(g)}},
\end{equation}   
where $z_{i,j}^{(g)}$ is the number of cyclic structures of order $g$
the edge $(i,j)$ belongs to, while $s_{i,j}^{(g)}$ is the number of
possible cyclic structures of order $g$ that can be built given the
degrees of the nodes.

We can now define, for every $g$, a detection algorithm that works
exactly as the GN method with the difference that, at every step, the
removed edges are those with the smallest value of ${\tilde C}_{i,j}^{(g)}$.
By considering increasing values of $g$, one can smoothly interpolate
between a local and a non local algorithm.  Notice that the definition
of ${\tilde C}_{i,j}^{(g)}$ guarantees that nodes with only one connection are
not considered as isolated communities by the algorithm, since for
their unique edge ${\tilde C}_{i,j}^{(g)}$ is infinite.

We have checked the accuracy of this new algorithm by comparing its
performance with the GN method.  Fig.~\ref{Fig4} reports the results
for the artificial test graph with four communities. It turns out that,
with respect to the strong definition, the new algorithm is as accurate
as the GN one, both in the case of cycles of order $g=3$
(triangles) and $g=4$ (squares).

On the other hand, for the weak definition
the best accuracy is achieved with the new algorithm with $g=4$.
Another test is performed by considering the examples of social
networks already studied by GN. Fig.~\ref{Fig5} shows the trees
resulting from the application of the GN algorithm and of the $g=4$
algorithm to the network of college football teams.  Again the results
are very similar, indicating that the local algorithm captures well
the presence of communities in that network.

Additional insight into the relationship between the GN algorithm and
this new algorithm based on edge clustering is provided by
Fig.~\ref{Fig6}, where the edge betweenness is plotted versus ${\tilde
C}_{i,j}^{(4)}$ for each edge of the graph of scientific
collaborations studied in Ref.~\cite{Newman01bis}.  It is clear that
an anticorrelation exists between the two quantities: edges with low
values of ${\tilde C}_{i,j}^{(4)}$ tend to have high values of
betweenness. The anticorrelation is not perfect: the edge with minimum
${\tilde C}_{i,j}^{(4)}$ is not the one with maximal
betweenness. Therefore we expect the two algorithms to yield similar
community structures though not perfectly coinciding.  

In this framework it is important to recall the parallel drawn by
Ronald S. Burt~\cite{burt} between betweenness centrality and the
so-called {\em redundancy}. The definition of redundancy is very close
to that of node clustering and, much in the spirit of our work, Burt
first pointed out that nodes that belong to few loops are central in
the betweenness sense.

Let us now turn our attention to the question of the computational
efficiency of our local algorithm. One can roughly estimate the
scaling of the computational time as follows. When an edge is removed
one has to check whether the whole system has been separated in
disconnected components and update the value of ${\tilde
C}_{i,j}^{(g)}$ in a small neighborhood of the removed edge.  The
first operation requires a time of the order of $M$, the total number
of edges present in the network, while the time required by the second
operation does not scale with
$M$. Since this operation has to be repeated for all edges we can
estimate the scaling of the total time as $a M + b M^2$. We thus
expect computational time to be linearly dependent on $M$ for
small systems and to cross over to an $M^2$ regime for large sizes.

We have measured the velocity of the new algorithm by computing the
time needed to generate the whole tree for a random graph of
increasing size $N$ and fixed average degree ($M \sim N$). Results,
reported in Fig.~\ref{Fig7}, confirm that the algorithm based on the
computation of the edge clustering coefficient is much faster
than the one based on edge betweenness, both for $g=3$ and $g=4$.  The
crossover between the initial linear dependence on $N$ to the $N^2$
growth for the local algorithm is evident for $g=3$.

\section{Community structure in a network of scientific collaborations}

In this section we consider an application of the fast algorithm
discussed in the previous section to a network of scientific
collaborations. In particular we have considered the network of
scientists who signed at least one paper submitted to the E-print
Archive relative to Condensed Matter in the period 1995-1999
({\tt http://xxx.lanl.gov/archive/cond-mat}).
Data have been kindly provided by Mark Newman.

The network includes $15616$ nodes (scientists) but it has a giant
connected component which includes only $N=12722$ scientists. We have
focused our attention on this giant component and we have applied to
it our algorithm for $g=3$ and $g=4$.  The time needed for the
generation of the dendrogram for $g=3$ is about 3 minutes on a desktop
computer with 800 MHz CPU.  The algorithm detects at the same time the
communities satisfying the weak and the strong criteria. At the end
of the procedure one obtains, for every value of $g$, a list of
communities identified in a weak sense and another list of communities
identified in a strong sense. By definition the second list is a
subset of the first one. Figure~\ref{dist_community} reports the size
distributions (for $g=3$ and $g=4$) of all the communities identified
in a weak sense. These distributions feature a power-law behaviour
$P(S) \sim S^{-\tau}$ with $\tau \simeq 2$, an indication of the
self-similar community structure of this network~\cite{guido_paolo}.
It is worth mentioning that the exponent we observe coincides with
the one obtained, using the GN algorithm, in the framework of a
recently proposed model of social network formation~\cite{boguna}.

For what concerns a more detailed analysis of the communities found,
validation of the results is far from trivial, since there is no
quantitative criterion to assess their accuracy. One may directly
inspect the dendrogram in order to answer questions like: are the
communities representative of real collaborations between the
corresponding scientists? Do they identify specific research areas?
Would a generic scientist agree about his or her belonging to a given
community?  Obviously all these questions cannot be answered in a
definitive and quantitative way. We have partially followed this path
and we have checked several subsets of the network at different levels
in the hierarchy. To the best of our knowledge the results seem to us
reasonable. Of course this does not represent a proof of the
efficiency of the algorithm. We refer the reader to the detailed
results of our analysis that we make available as additional
supporting information~\footnote{The list of all the communities
found with our algorithm for the scientific collaboration network
is available on request (castella@pil.phys.uniroma1.it)}.

\section{Conclusions}

The detection of the community structure in large complex networks is
a promising field of research with many open challenges.  The concept
of community is qualitatively intuitive.  However, to analyze a
network it is necessary to specify quantitatively and unambiguously
what a community is.  Once a definition is given it is in principle
possible to determine all subgraphs of a given network that fulfill
the definition.  However this task is in practice computationally out
of reach even for small systems.
Therefore the search for the community structure has generally a more limited
goal: selecting, among all possible communities, a subset of them organized
hierarchically, a dendrogram.  Divisive and agglomerative algorithms
carry out this task.  A comparison of the performances of such
algorithms is non trivial.  In some simple cases, as the artificial
graph with four subsets considered above, it is possible to assess
quantitatively the validity of the results.
In other cases, like the
network of scientific collaborations, no quantitative measure exists
to decide, given a precise definition of community, how good a
dendrogram is.  Typically, one may check whether the
results appear sensible.  However this is far from objective, being
mediated by the observer's own perception and by his/her intuitive concept of
community.

In this work we have proposed two improvements in the construction of
the dendrogram. First, we have devised a way to implement, in a
generic divisive algorithm, a quantitative definition of community.
In this way algorithms become fully self-contained, i.e. they do not
need non-topological input to generate the dendrogram.  Then we have
introduced a new divisive algorithm, which is based on local
quantities, and therefore extremely fast.  Both these improvements
have been tested successfully in controlled cases.  The analysis of
the large network of scientific collaborations gives results that
appear reasonable.  However, it is clear that, as discussed above,
this statement is subjective and cannot be made at present more
precise.  Definitely, a quantitative measure for the evaluation of
dendrograms would be a major step forward in this field.

At this point a remark is in order. So far we have only discussed
examples of the so-called social networks. It has been
shown~\cite{newman-soc-nonsoc} that social networks substantially
differ from other types of networks, namely technological or
biological networks. Among other differences, they exhibit a positive
correlation between the degree of adjacent vertices (assortativity)
while most non-social networks are disassortative. While these results
are consistent with our and other's findings about community
structures in social networks, they put into question the very
existence of a community structure in non-social networks and the
possibility of detecting it with the existing algorithms. From the
perspective of our local algorithm, which relies on the existence of
closed loops, disassortative networks could be in principle
problematic, due to the small number of short cycles. However,
interesting insight comes from the study of the loops of arbitrary
order~\cite{cal-pas-ves}. In particular for four different types of
networks (two of them social and assortative and two non-social and
disassortative) measured values for the so-called average grid
coefficient (the extension of the concept of clustering coefficient to
cycles of order four) are {\em two to four order of magnitude larger
than the corresponding coefficients of a random graph with the same
average degree and size $N$}. This argues in favor of the presence of
some sort of hierarchical structure and well-defined communities also
in disassortative networks. It also hints that our algorithm could be
fruitfully applied also to non-social (disassortative) networks,
although future work will be needed in this direction.

We believe that the new elements presented in this paper can be of
great help in the analysis of networks.  On the one hand, the
implementation of a quantitative definition of community makes
algorithms self-contained and allows the analysis of the community
structure based only on the network topology. On the other hand, the
introduction of a new class of local and fast algorithms could open
the way for applications to large-scale systems.

{\bf Acknowledgments:} The authors are very grateful to Mark Newman for
providing data on the networks of college football teams and of
scientific collaborations. They also wish to thank Alain Barrat for
useful suggestions and discussions.

\newpage

\centerline{\Huge Figure legends}

\vspace{1.0cm}

\begin{figure}
\includegraphics[angle=90,height=21.0cm]{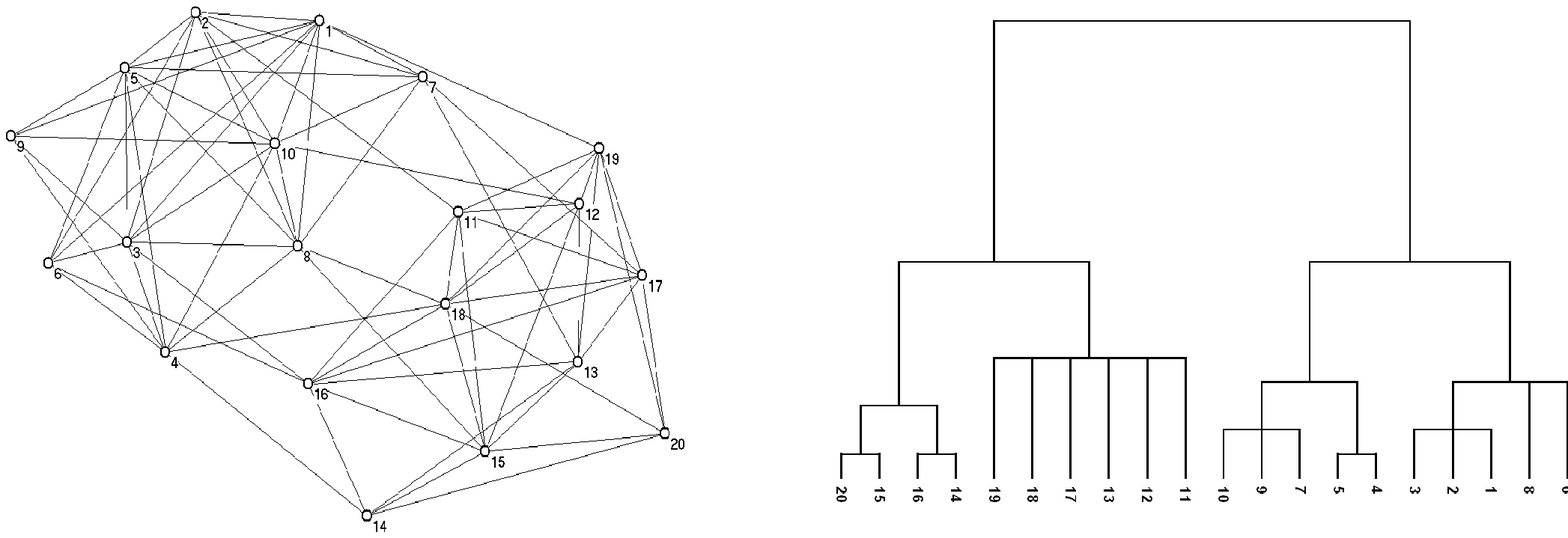}
\caption{}
\label{Fig1}
\end{figure}

\begin{figure}
\includegraphics[angle=90,height=21.0cm]{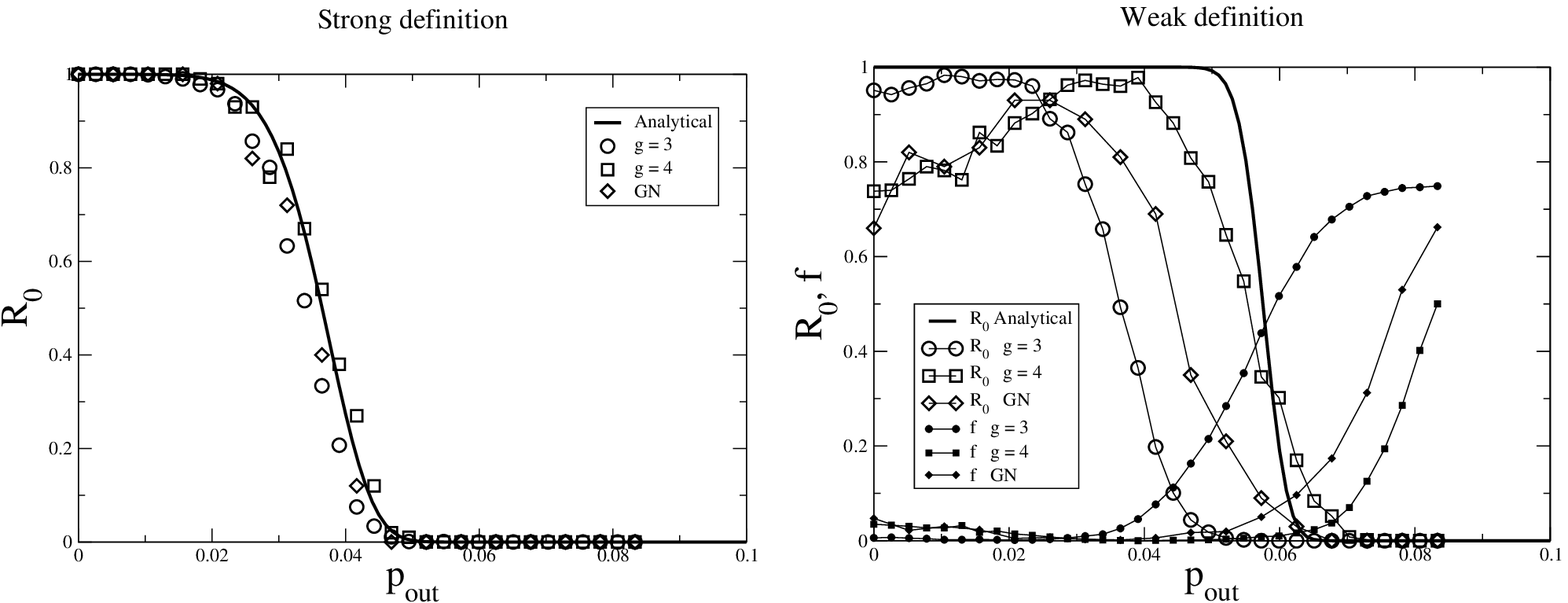}
\caption{}
\label{Fig4}
\end{figure}

\begin{figure}
\includegraphics[angle=90,height=21.0cm]{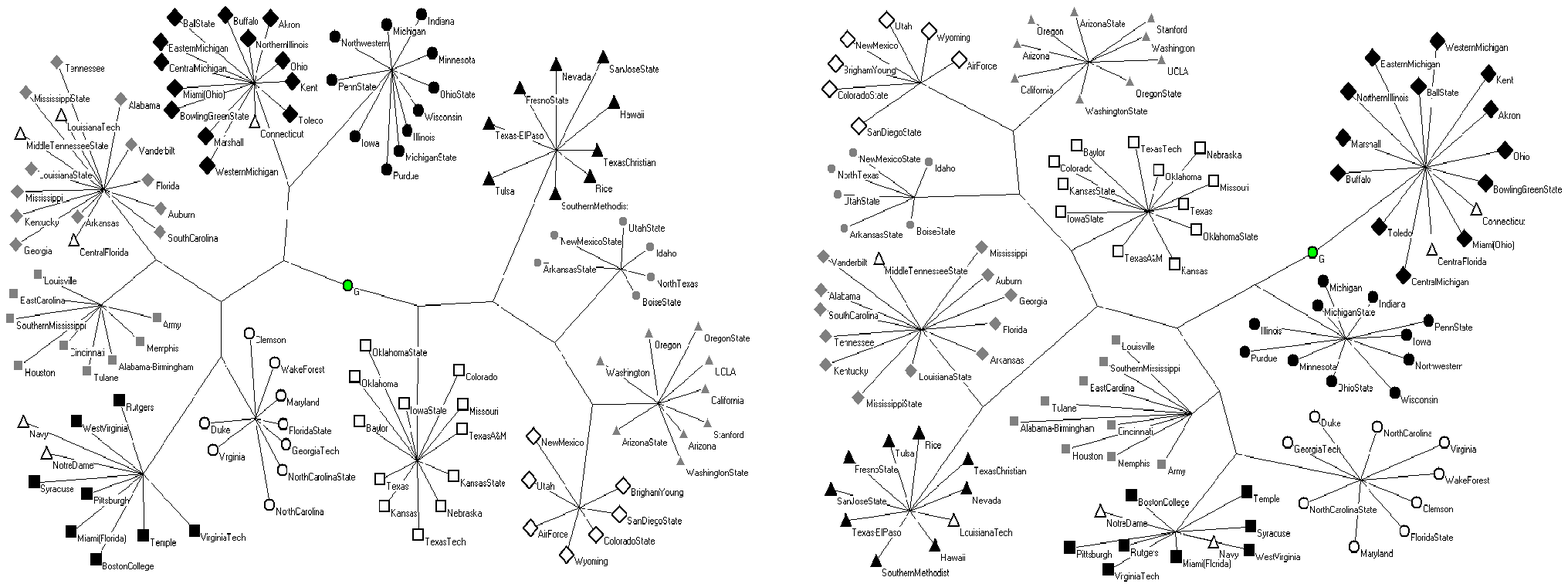}
\caption{}
\label{Fig5}
\end{figure}

\begin{itemize}

\item[Fig.1] A simple network (left) and the corresponding dendrogram
(right).

\item[Fig.2] Test of the efficiency of the different algorithms in the
analysis of the artificial graph with four communities. The
construction of the graph is described in text. Here $N=128$ and
$p_{in}$ is changed with $p_{out}$ in order to keep the average degree
equal to $16$. (Left) Strong definition: fraction of successes for the
different algorithms compared with the analytical probability that
four communities are actually defined.  (Right) Weak definition: in
addition to the same quantities plotted in the left graph, here we
report, for every algorithm, the fraction $f$ of nodes not correctly
classified.

\item[Fig.3] Plot of the dendrograms for the network of college football
teams, obtained using the Girvan-Newman algorithm (left) and our
new algorithm with $g=4$ (right). Different symbols denote teams
belonging to different conferences.  In both cases the observed
communities perfectly correspond to the conferences, with the
exception of the 6 members of the Independent conference, which are
misclassified.

\item[Fig.4] Edge betweenness vs. the modified edge clustering coefficient
${\tilde C}_{i,j}^{(4)}$, for the network of scientific collaborations
considered in section V. Each dot represents an edge in the
network. See section V for details.

\item[Fig.5] Plot of the average time (in seconds) needed to analyze a
random graph of $N$ nodes and fixed average degree $\langle k
\rangle=5$.  The time refers to the construction of the full tree down
to single nodes (Fig.~\ref{Fig1}). No criterion to validate the
communities was imposed. The runs are performed on a desktop computer
with a 800 MHz CPU.

\item[Fig.6] Normalized size distribution of all the communities of
scientists identified in a weak sense by the algorithm described in
section IV for $g=3$ (circles) and $g=4$ (squares). In both cases the
behaviour is well reproduced by a power law with exponent $-2$.

\end{itemize}

\newpage
\clearpage

\newpage
\clearpage

\newpage
\clearpage

\newpage
\clearpage

\begin{figure}
\vspace{0.3cm}
\includegraphics[angle=90,height=21.0cm]{Fig6.eps}
\caption{}
\label{Fig6}
\end{figure}

\newpage
\clearpage

\begin{figure}
\includegraphics[angle=90,height=21.0cm]{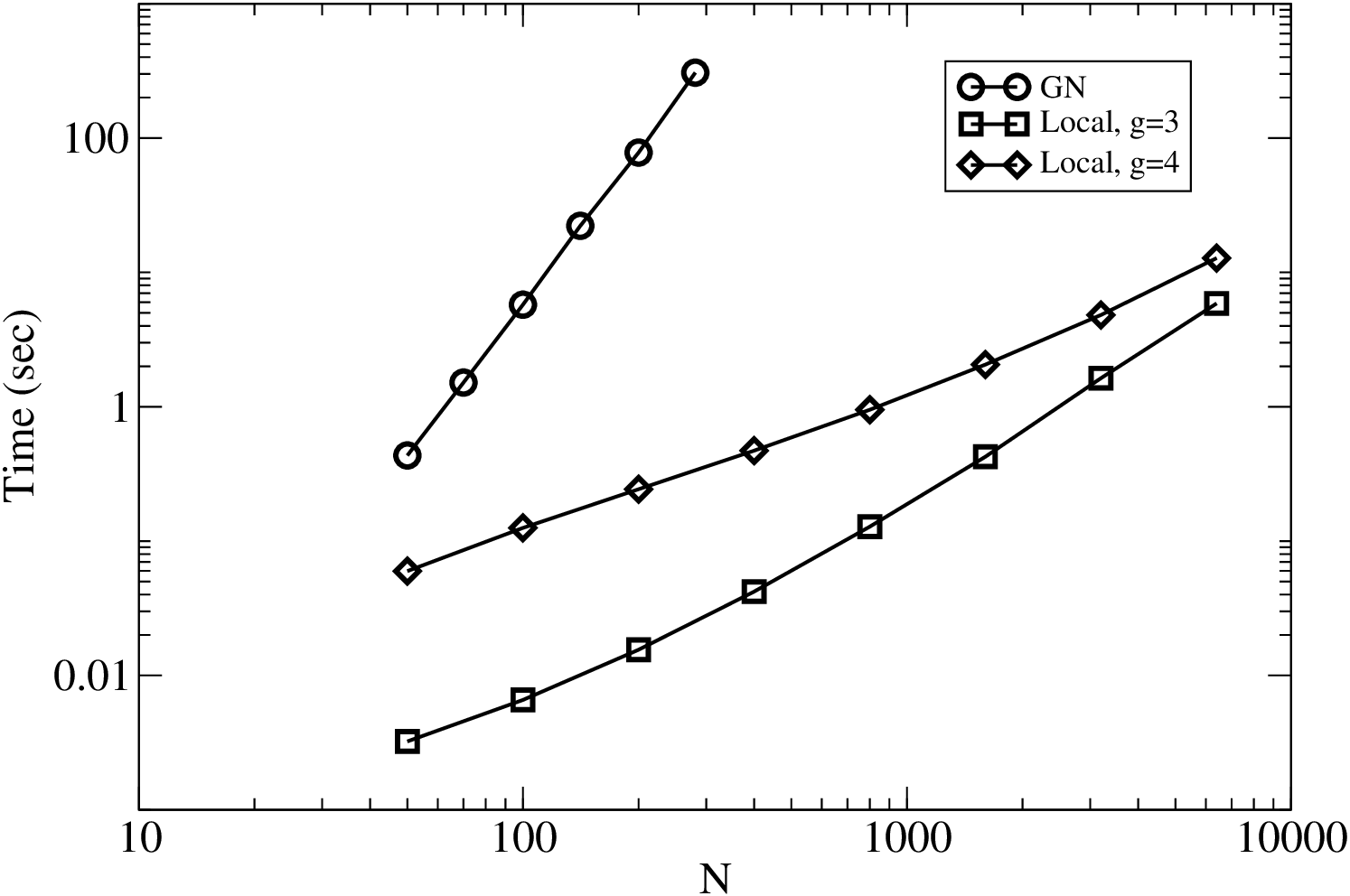}
\caption{}
\label{Fig7}
\end{figure}

\newpage
\clearpage

\begin{figure}
\includegraphics[angle=90,height=21.0cm]{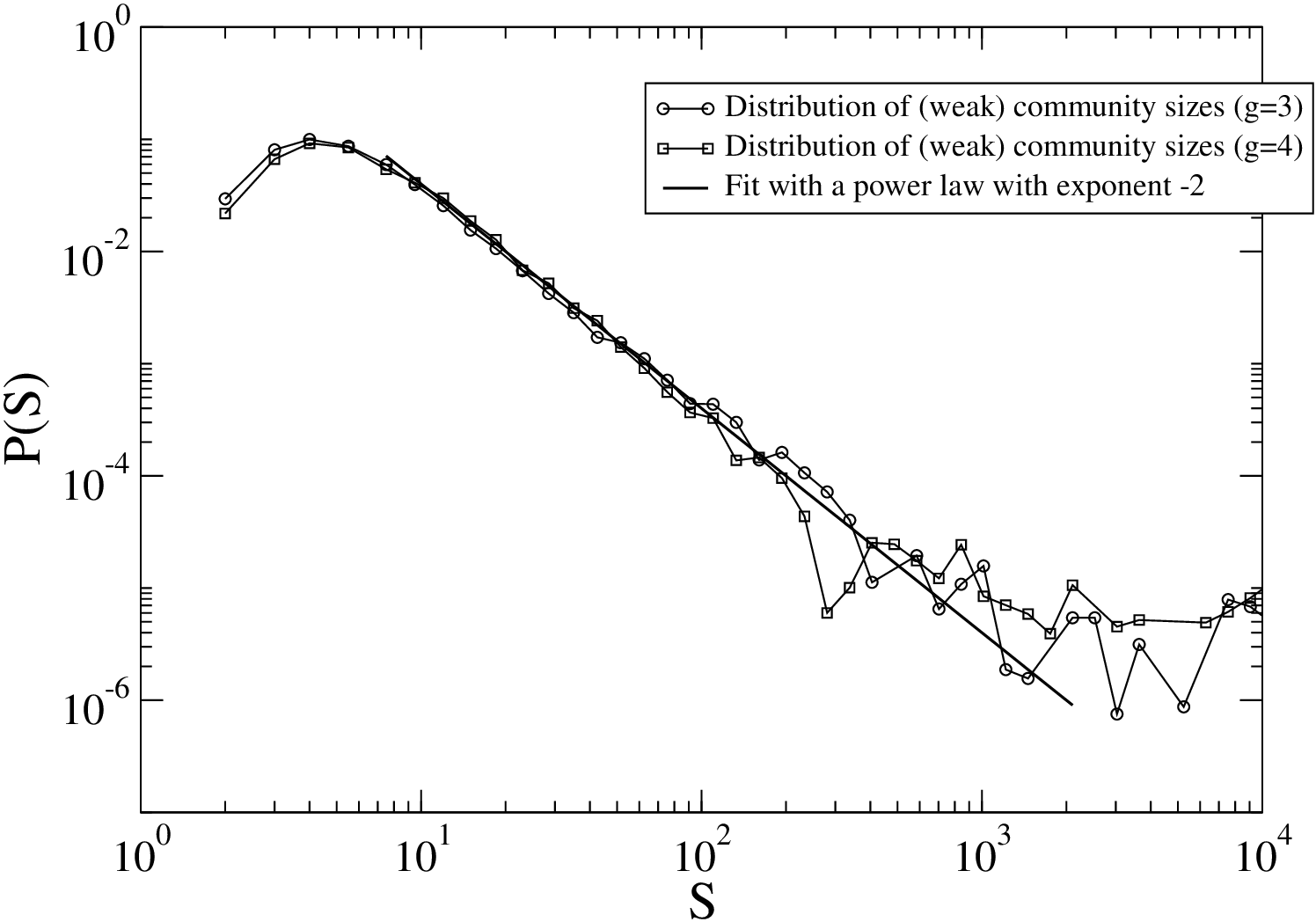}
\caption{}
\label{dist_community}
\end{figure}

\end{document}